\documentclass[runningheads]{llncs}
\usepackage{hyperref}

\usepackage{courier}
\usepackage{booktabs}
\usepackage{amsmath,amssymb}
\usepackage{algorithm2e}
\usepackage{listings}

\usepackage{tikz}
\usetikzlibrary{arrows,automata}
\usetikzlibrary{positioning}
\tikzset{initial text={},auto} 
\tikzstyle{every state}=[draw,shape=circle,inner sep=1pt,minimum size=10pt]

\lstdefinelanguage{mCRL2}
{
 keywords={act,var,cons,end,eqn,glob,init,val,whr,sort,map,pbes,proc,struct},
 keywords=[2]{true,false,delta,tau},
 keywords=[3]{Bool,Nat,Real,Pos,Int,Set,Bag,List,Int2Nat,Pos2Nat,Int2Pos,min,max
},
 keywords=[4]{hide,if,rename,sum,in,mu,nu,forall,exists,mod,allow,block,comm},
 keywords=[5]{nested,initial,state},
 numberstyle=\color{blue},
 comment=[l]\%,
 commentstyle=\slshape,
 keywordstyle=[1]\bfseries,
 keywordstyle=[2]\itshape,
 keywordstyle=[3]\itshape,
 keywordstyle=[4]\itshape,
 keywordstyle=[5]\bfseries\itshape,
 basicstyle=\ttfamily\scriptsize,
 flexiblecolumns=false,
 breaklines=false,
 tabsize=2,
 literate={\ \ }{{\ }}1
}
[keywords,comments]

\lstdefinelanguage{mCRL2-inline}
{
 keywords={act,var,cons,end,eqn,glob,init,val,whr,sort,map,pbes,proc,struct},
 keywords=[2]{true,false,delta,tau},
 keywords=[3]{Bool,Nat,Real,Pos,Int,Set,Bag,List,Int2Nat,Pos2Nat,Int2Pos,min,max
},
 keywords=[4]{hide,if,rename,sum,in,mu,nu,forall,exists,mod,allow,block,comm},
 keywords=[5]{nested,initial,state},
 numberstyle=\color{blue},
 comment=[l]\%,
 commentstyle=\slshape,
 keywordstyle=[1]\bfseries,
 keywordstyle=[2]\itshape,
 keywordstyle=[3]\itshape,
 keywordstyle=[4]\itshape,
 keywordstyle=[5]\bfseries\itshape,
 basicstyle=\ttfamily\footnotesize,
 flexiblecolumns=false,
 breaklines=true
}
[keywords,comments]

\newcommand{\inlinemcrl}{\lstinline[language=mCRL2-inline]}

\newcommand{\Bool}{\mathbb{B}}
\newcommand{\Nat}{\mathbb{N}}
\newcommand{\true}{\mathit{true}}
\newcommand{\false}{\mathit{false}}

\title{Tutorial: \\Designing Distributed Software in mCRL2}
\author{Jan Friso Groote\orcidID{0000-0003-2196-6587} \and Jeroen J.A. Keiren\orcidID{0000-0002-5772-9527}}
\institute{Department of Mathematics and Computer Science,\\Eindhoven University of Technology, The Netherlands\\
\email{\{J.F.Groote,J.J.A.Keiren\}@tue.nl}}
\begin{document}

\maketitle

\begin{abstract}
Distributed software is very tricky to implement correctly as some errors only occur in peculiar situations.
For such errors testing is not effective.
Mathematically proving correctness is hard and time consuming, and therefore, it is rarely done.
Fortunately, there is a technique in between, namely model checking, that, if applied with skill, is
both efficient and able to find rare errors.

\hspace*{0.3cm}In this tutorial we show how to create behavioural models of parallel software, how to specify
requirements using modal formulas, and how to verify these. For that we use the mCRL2 language and
toolset (\url{www.mcrl2.org/}).
We discuss the design of an evolution of well-known mutual exclusion protocols,
and how model checking not only
provides insight in their behaviour and correctness, but also guides their design.

\end{abstract}

\section{Introduction}
Whoever designed parallel or distributed software and protocols
must have found out how hard it is to get such software
correct.\footnote{In this paper, for the sake of brevity, we generally refer to parallel or distributed software just using the term distributed software. The techniques discussed in this paper apply equally in both situations.} Distributed software defies testing, as some errors only occur very rarely,
easily less than once in a million of runs. Yet, if such errors occur the software can go
awry, with effects that range from
confused internal administration, via crashing of the software, to loosing control over
safety-critical hardware.

The theoretical solution is to prove the correctness, for instance using assertional
methods that have been under development since the advent of the first electronic computers \cite{DBLP:journals/fac/AptO19}.
These days these methods are supported by proof checkers such as Coq
\cite{DBLP:series/txtcs/BertotC04} and Isabelle \cite{DBLP:books/sp/NipkowPW02}, or
integrated automatic provers for algorithms such as Dafny \cite{DBLP:journals/corr/LeinoW14}. These techniques are unprecedented
in locating software faults and are unbeatable if it comes to delivering correct software.
However, they have two important disadvantages. Proving the correctness of software can be very hard, as
the proof may require tricky combinatorial arguments, and detailed bookkeeping. More importantly, it is very
time consuming to provide a proof, even for a core algorithm, or a small distributed protocol.
The net result of this is that proving correctness of actual software is hardly ever used in
practical software development.

Fortunately, there is a method in between, namely model checking of models of the software.
The idea is to use an abstract modelling language to model the essence of the
distributed algorithm or protocol. Potential modelling languages with a powerful supporting model checking toolset
are mCRL2 \cite{DBLP:books/mit/GrooteM2014},
LNT \cite{DBLP:journals/sttt/GaravelLMS13} and FDR3 \cite{DBLP:journals/sttt/Gibson-Robinson16} as they support behaviour with parallelism as well as all commonly
used data types.
Standard programming languages such as Java and C++ are less suitable
for this purpose, as they are too versatile, and do not allow for concise mathematical formulation of
protocols and algorithms.
Domain Specific Languages to define automata based controllers such as ASD \cite{DBLP:journals/entcs/HopcroftB05} and Dezyne \cite{DBLP:conf/fmics/BeusekomGHHWWW17}
are suitable alternatives, with the advantage that they allow for code generation, but these languages
generally provide limited verification possibilities.

Only formulating models of distributed algorithms already substantially improves the quality of a subsequent implementation.
The reason is that models are more concise than
implementations, and models tend to be studied and discussed more thoroughly than programs.
Unfortunately, models still tend to contain errors and therefore, more needs to be done to increase
the quality.

Improving the quality of software models further can be done by providing alternative independent views on
the software and then comparing all views very precisely \cite{DBLP:journals/scp/BrandG15}.
The probability to make the same mistake in all views is the product of the probabilities
of making this mistake in each of the views. With a number of views the error probability drops dramatically,
and error probabilities of $10^{-10}$ are attainable.
In engineering such an approach is common where reliability is obtained due to redundancy.
Even checking light-weight properties can already make a substantial difference \cite{DBLP:journals/ese/OsaiweranSH14}.

We only take one alternative view, namely formulating compact properties on the model and proving them
using model checking. Other views are
making alternative models, independently making an implementation, specifying tests, and carrying out field tests.
The more alternative views, the higher the quality of the result, provided they are very precisely
compared to each other. Formulating correctness and proving this with
a proof checker is also a valid alternative view.

We use the modal mu-calculus with data as it is unsurpassed in expressivity \cite{DBLP:books/mit/GrooteM2014,DBLP:books/el/07/BradfieldS07}.
Fairness can be expressed using alternating fixed-points, and by using data
complex behaviour of the model can be tracked and analysed with modal formulas. Alternative property
languages, such as CTL/LTL can all be translated linearly to the modal mu-calculus with data
\cite{DBLP:journals/tcs/CranenGR11}.

In this tutorial we first describe mCRL2 and the modal mu-calculus very compactly. Subsequently, we
focus on mutual exclusion protocols for shared memory and traverse through the development of such protocols,
repeatedly identifying and repairing problems. We show how counterexamples that are
very helpful in identifying and understanding
problems \cite{DBLP:conf/cade/WesselinkW18}.
The modelling and analysis techniques described generalize to distributed algorithms in a straightforward way, using processes to model communication channels in stead of shared variables.
We thus hope that this tutorial will help
in understanding how to develop correct distributed algorithms and effectively obtain insight in
their behaviour, which goes far beyond showing that they terminate with the right response.

\section{mCRL2 primer}
In this section we give a concise description of mCRL2 and the modal mu-calculus.
More information is available in \cite{DBLP:books/mit/GrooteM2014}.
The language mCRL2 is based on process algebra \cite{DBLP:books/daglib/0067019,DBLP:conf/icalp/BergstraK84}.
The modal mu-calculus is based on Hennessy-Milner logic \cite{DBLP:journals/jacm/HennessyM85} and fixed point
equations \cite{DBLP:books/el/07/BradfieldS07}.

Process algebraic modelling centers around the notion of an action, typically denoted as \inlinemcrl{a},
\inlinemcrl{b}, \inlinemcrl{c}$\ldots$,
representing some atomic activity of a modelled entity, such as a program. Sending a message, writing a variable
or printing some text are typical examples. If actions must happen at exactly the same time we denote
them as multi-actions.
By writing \inlinemcrl{a|b} it is indicated that actions \inlinemcrl{a} and \inlinemcrl{b} happen at
the same instant in time. Actions and multi-actions have the same properties, and therefore we generally
only speak about actions in the sequel.

Using the sequential composition operator (\inlinemcrl{.}) actions
can be put in sequence and the choice operator ($+$) expresses that the behaviour of either the left
or the right operand can be done. A typical example
is \inlinemcrl{a.b+c.d} saying it is possible to do either an \inlinemcrl{a} followed by a \inlinemcrl{b}
or a \inlinemcrl{c} followed by a \inlinemcrl{d}.

Processes are specified by recursive equations.
The equation \inlinemcrl{proc P=a.P} indicates that the process \inlinemcrl{P} can infinitely often
do an \inlinemcrl{a} action. Using \inlinemcrl{init P} it is expressed that process \inlinemcrl{P} is
the behaviour defined by the specification.

Actions and processes can carry data, and all common data types are available. The process equation
\begin{center}
\inlinemcrl{proc Adder(n:Nat)=sum m:Nat.add(m).Adder(n+m)}
\end{center}
is an example.
The sum indicates the choice over all natural numbers.
This process can perform one of the actions \inlinemcrl{add(m)} for every number \inlinemcrl{m} and continues
with the behaviour \inlinemcrl{Adder(n+m)}. Behaviour can be executed conditionally on data
using the if-then-else operator \inlinemcrl{b->p<>q} where \inlinemcrl{b} is a boolean expression and
\inlinemcrl{p} and \inlinemcrl{q} are processes.

Processes are put in parallel using the parallel operator (\inlinemcrl{||}). Two parallel processes
can communicate by synchronising their actions. This is denoted using the communication
operator \inlinemcrl$comm({a_s|a_r->a},p||q)$, expressing that if action \inlinemcrl{a_s} and \inlinemcrl{a_r}
can happen in \inlinemcrl{p} resp.\ \inlinemcrl{q}, these actions can happen together as \inlinemcrl{a}.
We use the convention to write \inlinemcrl{_s} for send, and \inlinemcrl{_r} , after an action
if they will be used for a communication.
If actions \inlinemcrl{a_s} and \inlinemcrl{a_r} carry data they can only synchronise to \inlinemcrl{a} if
the data in both actions are equal. Then \inlinemcrl{a} will have this data as parameter as well. To enforce that
actions \inlinemcrl{a_s} and \inlinemcrl{a_r} must communicate, the allow operator is used.
The process \inlinemcrl$allow({a},p)$
expresses that only action \inlinemcrl{a} is allowed to happen in process \inlinemcrl{p} and all
other actions are blocked.

The modal mu-calculus is an extension
of propositional logic. Hence, we can use connectives such as \inlinemcrl{&&}, \inlinemcrl{||} and
\inlinemcrl{!} representing \textit{and}, \textit{or} and \textit{not}, respectively.
Writing \inlinemcrl{<a>phi}
expresses that an action \inlinemcrl{a} can be done after which \inlinemcrl{phi} holds, and
\inlinemcrl{[a]phi} expresses that if an action \inlinemcrl{a} is done, then \inlinemcrl{phi}
must hold afterwards. Instead of an action \inlinemcrl{a} we can use \inlinemcrl{true} to represent any action,
and \inlinemcrl{!a} to represent any action but \inlinemcrl{a}. We can use a Kleene star to indicate
arbitrary sequences of actions. So, \inlinemcrl{<!a*>phi} indicates that it is possible to do a sequence
of actions in which \inlinemcrl{a} does not occur such that afterwards \inlinemcrl{phi} holds. The formula
\inlinemcrl{[true*]phi} expresses that \inlinemcrl{phi} is valid after each sequence of actions.
All actions can carry data, and
quantification over data using \inlinemcrl{exists} and \inlinemcrl{forall} is possible.

Using the minimal fixed point operator \inlinemcrl{mu X.phi} and the maximal fixed point operator
\inlinemcrl{nu X.phi} recurring properties can be specified. By \inlinemcrl{nu X.<a>X} we express
that an infinite sequence of actions \inlinemcrl{a} must be possible. The formula \inlinemcrl{mu X.[!a]X&&<true>true}
says that the action \inlinemcrl{a} must be done on every path within a finite number of
actions. Using nested fixed points fairness properties can be expressed.

The fixed point variables can also use data. The following formula expresses that the total value offered to the adder will never exceed some maximum \inlinemcrl{M}:
\vspace{-1ex}
\begin{center}
\inlinemcrl{nu X(n:Nat=0).forall m:Nat.[add(m)]X(n+m) &&}\\
\inlinemcrl{             [!exists m:Nat.add(m)]X(n) &&}\\
\inlinemcrl{    val(n<M)}\phantom{PUSH LEFT}
\end{center}
\vspace{-1ex}
Here the variable \inlinemcrl{n}, initially equal to $0$, sums up all values of \inlinemcrl{m} occuring
in actions \inlinemcrl{add(m)}.
The box modality with the exists expresses that whenever an action different from \inlinemcrl{add} is done,
checking proceeds with an unaltered parameter \inlinemcrl{n}. Condition \inlinemcrl{n<M}
guarantees that the sum \inlinemcrl{n} never exceeds \inlinemcrl{M}. Keyword \inlinemcrl{val}
is needed to let the parser distinguish between modal formulas and data expressions.

\section{Mutual exclusion}

In this tutorial we study the mutual exclusion problem as we expect most of our readers to be
familiar with it. This allows us to focus on how the mCRL2 toolset helps us to
model and understand solutions for such a problem.
The techniques we describe are equally applicable in other problem domains.

Dijkstra describes the mutual exclusion problem as follows~\cite{Dij1965}:
\begin{quote}
``[\ldots] consider $N$ computers, each engaged in a process which, for our aims, can be regarded as cyclic.
In each of the cycles a so-called `critical section' occurs and the computers have to be programmed in such a way that at any moment only one of these $N$ cyclic processes is in its critical section.''
\end{quote}

The first solution to the mutual exclusion problem has been known since 1959.
It was first described by Dijkstra~\cite{Dij1962}, who attributed it do Dekker.
In this paper Dijkstra also show two simpler incorrect solutions.
A first solution for $N$ processes is due to Dijkstra~\cite{Dij1965} and only much
later the well-known solution by Peterson appeared~\cite{Pet1981}.

From Section \ref{sec:first} onward we model Dijkstra's algorithms in increasing complexity.
Subsequently, we investigate Peterson's mutual exclusion algorithm.

\paragraph{Requirements.}
Before modelling solutions, we ask ourselves what the properties are that a mutual exclusion
protocol should have. In order to understand the requirements it is necessary to understand that
we model mutual exclusion using three phases. First, a \textit{wish} is indicated to enter the
critical section, second access is granted indicated by \textit{enter}, after which the process
indicates that is left the critical section using \textit{leave}.

\begin{description}
  \item[Mutual exclusion.] At any moment only one of the processes is in its critical section.
  \item[Always eventually request.] Every process can always eventually wish to enter its critical section.
  \item[Eventual access.] Whenever a process indicates a wish to enter is critical section, it is guaranteed
       to eventually get access to its critical section. This property is also referred to as starvation freedom.
  \item[Bounded overtaking.] There is an absolute bound $B$ such that, whenever a process indicates it wants to
        enter its critical section, at most $B$ processes can enter their critical section, before this
        process enters its critical section.
\end{description}
It is natural to formulate mutual exclusion as a property. But mutual exclusion is insufficient,
as it can easily be guaranteed by never letting a process enter the critical section. For a properly
functioning mutual exclusion protocol the second and third properties are equally important. The last one is interesting
especially in systems where execution of programs does not need to be fair.

\paragraph{Memory model.}
We assume that the mutual exclusion protocols are implemented on a platform with shared memory where
variables are written and read consecutively in some interleaved fashion by the parallel programs.

%
%
%
%

\subsection{A naive algorithm for mutual exclusion}
\label{sec:first}
For two processes a naive solution of the mutual exclusion problem is Algorithm~1 suggested by
Dijkstra~\cite{Dij1962}.
It uses two global Boolean variables $\mathit{flag}[i]$ in which process $i$ indicates that it is in its
critical section.
The algorithm, for process $i$ now proceeds as follows.
It first blocks until the flag of process $1{-}i$ becomes $\false$ using busy waiting.
Once $\mathit{flag}[1{-}i]$ is $\false$, the other process is not in its critical section.
It then sets its own flag to $\true$ and enters its critical section.
Once the work in the critical section is complete, it sets its flag to $\false$.
\vspace{-1ex}
\begin{center}
\begin{tabular}{l}
\textbf{Data:} Global variables $\mathit{flag}[0], \mathit{flag}[1] \colon \Bool$\\
\textbf{while}~{$\mathit{flag}[1{-}i]$} \textbf{do} \texttt{/* Busy waiting */}~\textbf{end}\\
$\mathit{flag}[i] := \true$;\\
  \texttt{/* Critical section */}\\
$\mathit{flag}[i] := \false$;
\end{tabular}\vspace{1ex}\\
{\small \textbf{Algorithm 1:} A naive mutual exclusion algorithm for process $i$.}
\label{alg:mutex-naive}

\end{center}\vspace{-1ex}
Below we go through a few steps to model this algorithm in mCRL2.

\paragraph{Shared variables.}
A shared variable can be modelled as process that carries the current value of the variable as a parameter.
It can perform a read action, in which it sends the current value of its parameter.
Also, it can perform a write action for each possible value that can be stored in the variable.
The array $\mathit{flag}$ is modelled by the following process.

\begin{lstlisting}[language=mCRL2,linerange=10-12]
 proc Flag(i:Nat, b:Bool)=
               sum b':Bool. set_flag_r(i, b').Flag(i, b') +
               get_flag_s(i, b).Flag(i, b);
\end{lstlisting}
The name of the process is \inlinemcrl{Flag}. The parameter \inlinemcrl{i:Nat} describes the index in the array,
and \inlinemcrl{b:Bool} gives the current value of the variable.
Using \inlinemcrl{sum b':Bool.set_flag_r(i,b').Flag(i,b')} we model that the process can receive any new value \inlinemcrl{b'} from another process, and store it to parameter \inlinemcrl{b}.
The action \inlinemcrl{get_flag_s(i, b).Flag(i, b)} allows to send the current value to any process that
requests the value.

\paragraph{Modelling the busy waiting loop.}
The effect of the busy waiting loop is that the process can only continue when the guard becomes $\false$, i.e., when $\mathit{flag}[1{-}i]$ has value $\false$.
We could model the busy waiting loop explicitly by a recursive process.
However, in mCRL2, a read action blocks until the matching send action can also be performed.
As the shared variable only sends its current value, we can model this loop by using \inlinemcrl{get_flag_r(1-i, false)}, that is, reading $\false$ from the flag of the other process.
Since a subtraction results in an integer instead of a natural number, we need to add an explicit
type conversion here, and write \inlinemcrl{get_flag_r(Int2Nat(1-i), false)}.
As \inlinemcrl{i} is either 0 or 1, this is guaranteed to be natural number.

\paragraph{Model.}

We now combine this into an mCRL2 model.
First, we define the behaviour of process $i$.

\begin{lstlisting}[language=mCRL2,linerange=14-20]
proc  Mutex(i:Nat) = 
               get_flag_r(Int2Nat(1-i), false).
               set_flag_s(i, true).
               enter(i).
               leave(i).
               set_flag_s(i, false).
               Mutex();
\end{lstlisting}
Note that the process is the sequential composition of the busy waiting loop,
setting the flag of process $i$ to $\true$ using \inlinemcrl{set_flag_s(i, true)},
entering the critical section using action \inlinemcrl{enter(i)}, leaving it using
\inlinemcrl{leave(i)}, and setting the flag to $\false$ again.
At the end of the algorithm we write \inlinemcrl{Mutex()} to model
that the critical section can repeatedly be entered.
Writing \inlinemcrl{Mutex()} without parameters is a shorthand that leaves
the current value of the parameters unchanged.
Here it is thus equivalent to writing \inlinemcrl{Mutex(i)}.

The system as a whole consists of two instances of \inlinemcrl{Mutex} and two shared variables,
synchronising on \inlinemcrl{get_flag} and \inlinemcrl{set_flag}.

\begin{lstlisting}[language=mCRL2,firstline=22]
init allow({enter, leave, get_flag, set_flag},
             comm({get_flag_r | get_flag_s -> get_flag,
                          set_flag_r | set_flag_s -> set_flag},
               Mutex(0) || Mutex(1) || Flag(0,false) || Flag(1,false)));
\end{lstlisting}
Here, the operator \inlinemcrl!comm! specifies that, \inlinemcrl{get_flag_r} and \inlinemcrl{get_flag_s}
can synchronise.
The result is is named \inlinemcrl{get_flag}.
It does the same for \inlinemcrl{set_flag_r} and \inlinemcrl{set_flag_s}.
Writing \inlinemcrl!allow{enter, leave, get_flag, set_flag}! specifies that we are only interested
in the result of the communication, essentially enforcing synchronisation. We also allow the actions
\inlinemcrl{enter} and \inlinemcrl{leave} that
are local to the processes, and hence do not participate in any synchronisation.

\paragraph{Verification.}
Now that we have a model of this first mutual exclusion algorithm, we focus on its correctness.
How can we formalize the mutual exclusion property using the mu-calculus?
Observe that we explicitly modelled entering and leaving the critical section.
Process $i$ is therefore in its critical section if it performed an
\inlinemcrl{enter(i)} action, but has not yet done the corresponding \inlinemcrl{leave(i)}.
Mutual exclusion is then violated if we see two \inlinemcrl{enter} actions without an
intermediate \inlinemcrl{leave}.
This is captured in the following mu-calculus formula.

\begin{lstlisting}[language=mCRL2]
[true*][exists i1:Nat.enter(i1)]
             [!(exists i2:Nat.leave(i2))*][exists i3:Nat.enter(i3)]false
\end{lstlisting}

This formula expresses that invariantly (\inlinemcrl{[true*]}), after a process enters its critical section (\inlinemcrl{[exists i1:Nat.enter(i1)]}), as long as no leave action happened
(\inlinemcrl{[!(exists i2:Nat.leave(i2))*]}), another process is not allowed to enter its critical section (\inlinemcrl{[exists i3:Nat.enter(i3)]false}).

\begin{figure}[!ht]
  \vspace{-4ex}
  \begin{center}
    \hspace*{-0.5cm}\includegraphics[width=1.1\textwidth]{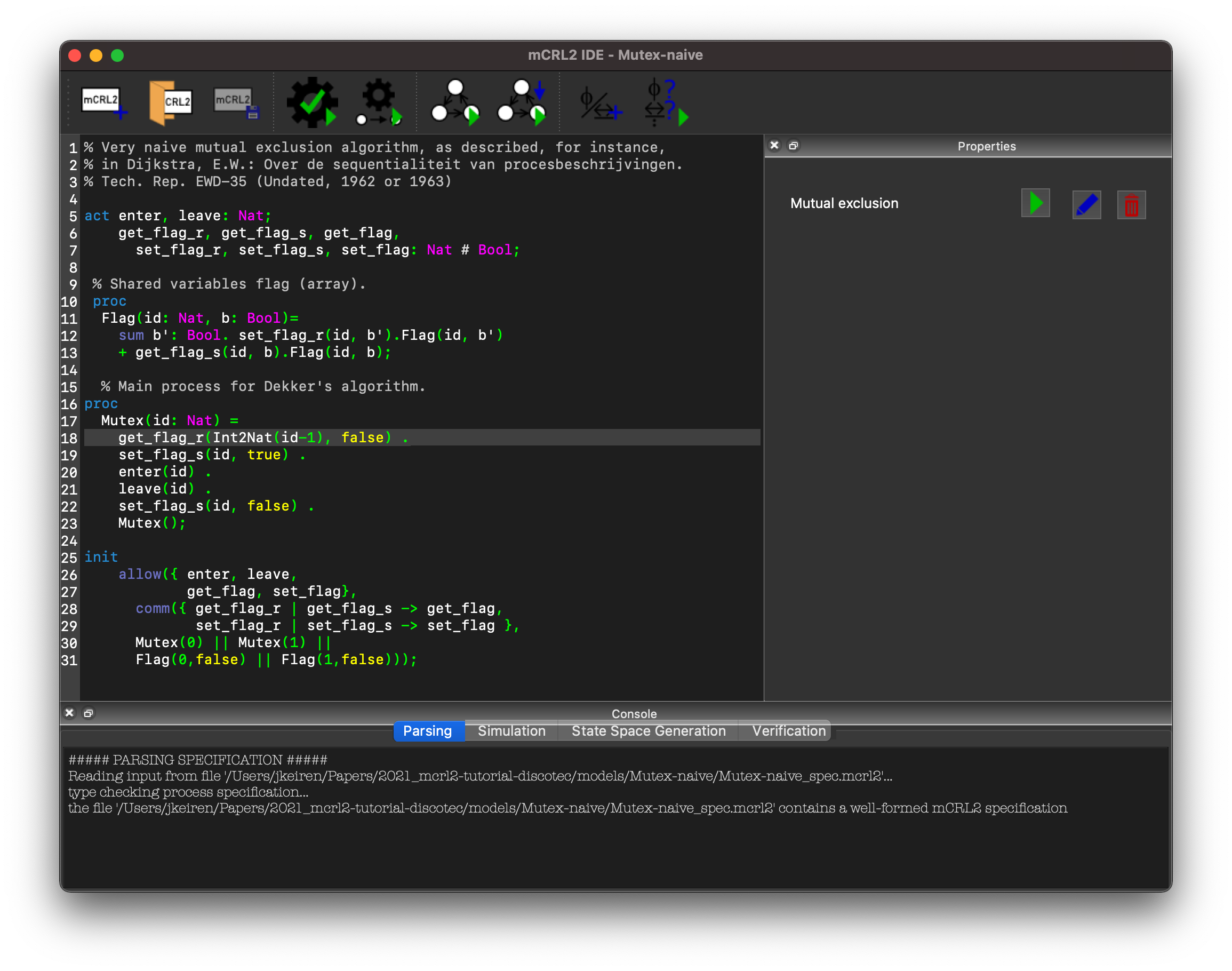}
  \vspace{-11ex}\\
  \end{center}
    \caption{Screenshot of mCRL2ide with naive mutual exclusion algorithm.}
    \label{fig:mutex-naive-ide}
\end{figure}

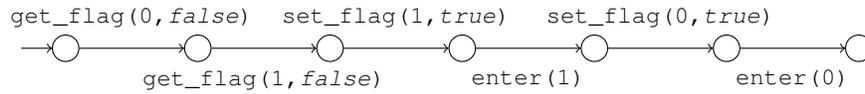
\begin{figure}[!ht]
  \centering
  \begin{tikzpicture}[node distance=50pt]
    \lstset{language=mCRL2}
    \node[state, initial left] (s0) {};
    \node[state, right of=s0] (s1) {};
    \node[state, right of=s1] (s2) {};
    \node[state, right of=s2] (s3) {};
    \node[state, right of=s3] (s4) {};
    \node[state, right of=s4] (s5) {};
    \node[state, right of=s5] (s6) {};

    \draw[->]
      (s0) edge node[above,yshift=5pt] {\inlinemcrl{get_flag(0,false)}} (s1)
      (s1) edge node[below,yshift=-5pt] {\inlinemcrl{get_flag(1,false)}} (s2)
      (s2) edge node[above,yshift=5pt] {\inlinemcrl{set_flag(1,true)}} (s3)
      (s3) edge node[below,yshift=-5pt] {\inlinemcrl{enter(1)}} (s4)
      (s4) edge node[above,yshift=5pt] {\inlinemcrl{set_flag(0,true)}} (s5)
      (s5) edge node[below,yshift=-5pt] {\inlinemcrl{enter(0)}} (s6)
      ;
  \end{tikzpicture}
  \vspace{-2ex}
  \caption{Counterexample of the mutual exclusion property for the naive algorithm.}
  \label{fig:mutex-naive-counterexample}
\end{figure}

We entered the model and the property in \texttt{mcrl2ide}, which is mCRL2's IDE that supports most basic
uses of the mCRL2 toolset.
A screenshot is shown in Figure~\ref{fig:mutex-naive-ide}. By clicking the `Verify' button (green triangle)
of the mutual exclusion property, the tools will verify whether the property holds.
In this case, it finds that the property is violated, and the `Verify' button changes into a red `C'. By
clicking the red `C', the tool shows a counterexample. In this case, the counterexample is the one shown
in Figure~\ref{fig:mutex-naive-counterexample}.

This counterexample is a trace where two processes execute an
\inlinemcrl{enter}\ action without in intermediate \inlinemcrl{leave}.
If we check the counterexample, it is immediately clear what is going on.
Both processes check simultaneously that the other process is not in its critical section, concluding
they can proceed to their critical section.
Then, process $1$ sets its flag and enters its critical section, immediately followed by process $0$.

\subsection{Fixing the naive algorithm}

The problem with the naive algorithm is that each process first checks if the other process is in
its critical section, and then sets its own flag.
If this is done simultaneously, the processes do not observe that the other process is entering
the critical section at the same time.
We could potentially resolve this issue by first setting the flag, expressing the intent to enter
the critical section, and then only proceed into the critical section if the
flag of the other process is false.
The improved algorithm is shown in Algorithm~2. It also stems from~\cite[Fig. 2]{Dij1962}.

\begin{center}
\begin{tabular}{l}
\textbf{Data:} Global variables $\mathit{flag}[0], \mathit{flag}[1] \colon \Bool$\\
$\mathit{flag}[i] := \true$;\\
\textbf{while} $\mathit{flag}[1{-}i]$ \textbf{do} \texttt{/* Busy waiting */} \textbf{end}\\
\texttt{/* Critical section */}\\
$\mathit{flag}[i] := \false$;
\end{tabular}\vspace{2ex}\\
{\small \textbf{Algorithm 2:} Improved naive mutual exclusion algorithm for process $i$.}
\end{center}

\paragraph{Model.}
The change in the mCRL2 model is equally simple. We only exchange the first two lines of the \inlinemcrl{Mutex} process, which now becomes the following.

\begin{lstlisting}[language=mCRL2,linerange=17-24]
proc Mutex(i:Nat) = 
         set_flag_s(i, true).
         get_flag_r(Int2Nat(1-i), false).
         enter(i).
         leave(i).
         set_flag_s(i, false).
         Mutex();
\end{lstlisting}

\paragraph{Verification.}
Changing the order of the program fixed the algorithm as it now satisfies the mutual exclusion property.
So, we investigate the requirement that every process can always
eventually wish to enter its critical section.

If process $i$ sets its flag to $\true$ this means that it expresses the wish to enter its critical section.
The property can then be expressed by saying that invariantly, for all processes $i$ there is a path to a state in which process $i$ can set its flag to $\true$.
This is expressed in the mu-calculus as follows.

\begin{lstlisting}[language=mCRL2]
[true*]forall i:Nat.val(i<=1) => <true*><set_flag(i, true)>true
\end{lstlisting}

Recall that \inlinemcrl{[true*]phi} is valid if \inlinemcrl{phi} holds in all reachable states.
We express the property for all processes $i$ using \inlinemcrl{forall i:Nat}
with \inlinemcrl{val(i<=1)}.
The remaining formula \inlinemcrl{<true*><set_flag(i, true)>true} expresses that there is a path to a state in
which \inlinemcrl{set_flag(i, true)} can happen.

\begin{figure}[!ht]
  \centering
  \begin{tikzpicture}[node distance=100pt]
    \lstset{language=mCRL2}
    \node[state, initial left] (s0) {};
    \node[state, right of=s0] (s1) {};
    \node[state, right of=s1] (s2) {};

    \draw[->]
      (s0) edge node[above] {\inlinemcrl{set_flag(1,true)}} (s1)
      (s1) edge node[above] {\inlinemcrl{set_flag(0,true)}} (s2)
      ;
  \end{tikzpicture}
  \caption{Counterexample showing that a process cannot always eventually wish to enter.}
  \label{fig:fixed-mutex-naive-counterexample}
\end{figure}
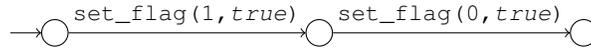

When we verify this property, it turns out that it does not hold.
The counterexample is shown in Figure~\ref{fig:fixed-mutex-naive-counterexample}.

\vspace{-4ex}
This counterexample shows that if the processes simultaneously wish to enter the critical section,
they block each other's possibility to proceed, disallowing both processes to set their own flag to
$\false$, leading to a deadlock. Due to the deadlock, they can never express their wish to reenter the
critical section.

\subsection{Dekker's algorithm}

To resolve the deadlock in the previous mutual exclusion algorithm, Dekker's solution is to give priority to one of the two processes whenever both processes want to enter their critical section.
We present it as Algorithm~3~\cite{Dij1962}.

\begin{center}
\begin{tabular}{l}
  \textbf{Data:} Global variables $\mathit{flag}[0], \mathit{flag}[1] \colon \Bool$ and
              $\mathit{turn} \colon \Nat$\\
  $\mathit{flag}[i] := \true$;\\
  \textbf{while} $\mathit{flag}[1{-}i]$ \textbf{do}\\
  \hspace*{0.5cm}\textbf{if} $\mathit{turn} \neq i$ \textbf{then}\\
  \hspace*{1cm}$\mathit{flag}[i] := \false$;\\
  \hspace*{1cm}\textbf{while} $\mathit{flag}[1{-}i]$ \textbf{do}
        \texttt{/* Busy waiting */} \textbf{end}\\
  \hspace*{1cm}$\mathit{flag}[i] := \true$;\\
  \hspace{0.5cm}\textbf{end}\\
  \textbf{end}\\
  $\mathit{turn} := 1{-}i$;\\
  \texttt{/* Critical section */}\\
  $\mathit{flag}[i] := \false$;
\end{tabular}\vspace{2ex}\\
{\small \textbf{Algorithm 3:} Dekker's algorithm for process $i$.}
\end{center}
The idea behind the algorithm is as follows.
First, compared to the previous attempts, the meaning of global Boolean variables $\mathit{flag}[i]$ changes, and now indicates whether process $i$ wishes to access its critical section.
Second, a new shared variable $\mathit{turn}$ indicates which process has priority when both processes want to enter their critical section.\footnote{In \cite{Dij1962}, variables $\mathit{LA}$ and $\mathit{LB}$ are used as flags, and a Boolean variable $\mathit{AP}$ is used in the place of $\mathit{turn}$.}
The key idea now is that, while the other process $1{-}i$ wishes to enter its critical section, process $i$ checks
whether the other process has priority. If so, process $i$ sets its flag to $\false$, and then waits until
process $1{-}i$ leaves its critical section and sets its flag to $\false$.
Then process $i$ resets its flag to $\true$ and continues as before.

\paragraph{Model.}

To model this algorithm in mCRL2, we have to decide how to deal with the outer loop and the if-clause. We first discuss how to model the outer while-loop.

In more general terms, we want to model a program
$S_1; \mathbf{while}\ b\ \mathbf{do}\ S_2\ \mathbf{end}; S_3$ in mCRL2.
The most straightforward way to model this is to have two separate processes that are executed sequentially.
The first process performs the behaviour of $S_1$ and hands execution over to the second process.
The second process evaluates $b$.
If $b$ is $\true$ it executes the behaviour of $S_2$ and then executes itself, repeating the behaviour.
Otherwise it executes the behaviour of $S_3$.

An if-then clause  \textbf{if} $b$ \textbf{then} $S_1$ \textbf{end}; $S_2$ can be modelled directly into the if-then-else construct \inlinemcrl{b->p<>q} of mCRL2. In this case \inlinemcrl{p} is the translation of $S_1;S_2$ and
\inlinemcrl{q} is the translation of $S_2$.
Note that both for the loop and the if-then clause, if the condition contains shared variables, their values must first be read.

The outer loop of the algorithm is modelled as follows.

\begin{lstlisting}[language=mCRL2,linerange=32-47]
proc Dekker_outer_loop(i:Nat) = 
	            sum flag_other:Bool.get_flag_r(other(i), flag_other). 
	            flag_other -> (sum turn: Nat.get_turn_r(turn).
                                           (turn != i) -> (set_flag_s(i, false).
                                                                           get_turn_r(i).      
                                                                           set_flag_s(i, true). 
                                                                           Dekker_outer_loop(i) 
                                                                          ) 
                                                                   <> Dekker_outer_loop(i)    
                                            ) 
                                      <> (set_turn_s(other(i)).   
                                             enter(i).
                                             leave(i).
                                             set_flag_s(i, false).   
                                             Dekker(i);
                                            );
\end{lstlisting}

Note that the shared variable \inlinemcrl{flag} of the other process is read, and its value is stored in \inlinemcrl{flag_other}.
If the guard of the outer loop is true (\inlinemcrl{flag_other ->}),  the loop is entered.
In the body of the loop the turn variable is read, and it is decided whether the if-clause must be entered.
In the body of the if the flag for this process is set to false, allowing the other process to enter its critical section. The process waits until the other process leaves its critical section.
Note that here we use the construct we previously introduced for the busy waiting loop.
Subsequently, the flag of this process is set to true, and the while loop is repeated.

If the guard of the outer loop is false, we jump to the else part starting with the lower
\inlinemcrl{<>} symbol, from where the rest of the process similar to our previous algorithms.

For the complete model, we also need a process modelling the global variable $\mathit{turn}$. This is done in
a similar way as for the global array $\mathit{flag}$, where the variable is set and read using
actions \inlinemcrl{set_turn} and \inlinemcrl{get_turn}, which are the results of synchronising
\inlinemcrl{set_turn_r} and \inlinemcrl{set_turn_s}, and \inlinemcrl{get_turn_s} and
\inlinemcrl{get_turn_r}.\linebreak[3]
The parallel composition must be extended with the process modelling $\mathit{turn}$, as well as with
an increased number of synchronising actions, and is given below. Some aspects of this process expression are
explained in the next part on verification.

\begin{lstlisting}[language=mCRL2,linerange=49-55]
init allow({wish|set_flag, enter, leave,
                        get_flag, set_flag, get_turn, set_turn},
             comm({get_flag_r | get_flag_s -> get_flag,
                          set_flag_r | set_flag_s -> set_flag,
                          get_turn_r | get_turn_s -> get_turn,
                          set_turn_r | set_turn_s -> set_turn},
            Dekker(0) || Dekker(1) || Flag(0,false) || Flag(1,false) || Turn(0)));
\end{lstlisting}

\paragraph{Verification.}
As the algorithm keeps the same logic guarding the critical section as before,  mutual exclusion is still satisfied.
This is easily verified using modal formula given earlier.

However, to verify that we can always eventually request access to the critical section
we need to be more careful.
So far, we assumed that when a process sets its flag, this corresponds to expressing the wish to enter the critical section.
However, as in Dekker's algorithm there are multiple places where the flag is set to true, we do not have this nice one-to-one correspondence.
We therefore amend the model with an action \inlinemcrl{wish(i)} that makes the wish explicit the first time the process sets its flag.
The main process therefore becomes the following.

\begin{lstlisting}[linerange=28-30,language=mCRL2]
proc Dekker(i:Nat) = 
                wish(i)|set_flag_s(i, true).
	            Dekker_outer_loop(i);
\end{lstlisting}

We here use a multi-action to model that \inlinemcrl{wish} and \inlinemcrl{set_flag} happen simultaneously.
The set of allowed actions needs to be extended with \inlinemcrl{wish|set_flag}.
We also need to modify the property to check for a such a multi-action instead of just the \inlinemcrl{set_flag}, hence the formula  for always eventual request becomes the following.

\begin{lstlisting}[language=mCRL2]
[true*]forall i:Nat.val(i<=1) => <true*><wish(i)|set_flag(i, true)>true
\end{lstlisting}
This formula holds for Dekker's algorithm.

We now look at the property of eventual access.
This says that, whenever a process wishes to enter its critical section, it inevitably ends up in the critical section.
This can be formulated using the following mu-calculus formula.

\begin{lstlisting}[language=mCRL2]
[true*]forall i:Nat.val(i<=1) 
           => [exists b:Bool.wish(i)|set_flag(i,b)]mu X.([!enter(i)]X && <true>true)
\end{lstlisting}

\begin{figure}[!ht]
\centering
\resizebox{\textwidth}{!}{
  \begin{tikzpicture}[node distance=100pt]
    \lstset{language=mCRL2}
    \node[state, initial left] (s0) {};
    \node[state, right of=s0] (s1) {};
    \node[state, right of=s1] (s2) {};
    \node[state, right of=s2] (s3) {};
    \node[state, right of=s3] (s4) {};
    \node[state, right of=s4] (s5) {};

    \draw[->]
      (s0) edge node[above,yshift=5pt] {\inlinemcrl{wish(0)|set_flag(0,true)}} (s1)
      (s1) edge node[below,yshift=-5pt] {\inlinemcrl{get_flag(1,false)}} (s2)
      (s2) edge node[above,yshift=5pt] {\inlinemcrl{set_turn(1)}} (s3)
      (s3) edge node[below,yshift=-5pt] {\inlinemcrl{wish(1)|set_flag(1,true)}} (s4)
      (s4) edge[bend left] node[above,yshift=5pt] {\inlinemcrl{get_flag(0,true)}} (s5)
      (s5) edge[bend left] node[below,yshift=-5pt] {\inlinemcrl{get_turn(1)}} (s4)
      ;
  \end{tikzpicture}}
  \caption{Counterexample of the eventual access property for Dekker's algorithm.}
  \label{fig:dekker-eventual-access-counterexample}
\end{figure}
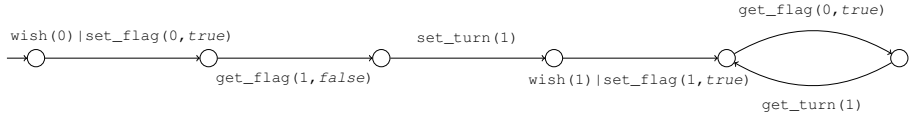

The formula says that invariantly, for every valid process \inlinemcrl{i}, when \inlinemcrl{i} wishes to enter its critical section (\inlinemcrl{[exists b:Bool.wish(i)|set_flag(i,b)]}), an \inlinemcrl{enter(i)} action inevitably happens within a finite number of steps (\inlinemcrl{mu X.([!enter(i)]X && <true>true)}).
The conjunction \inlinemcrl{<true>true} ensures that that last part of the formula does not hold trivially in a deadlock state.
Verifying the property yields false, and we get the counterexample shown in Figure~\ref{fig:dekker-eventual-access-counterexample}.

The counterexample is interesting.
It describes the scenario where process 0 requests access to its critical section, by setting the flag.
It then checks the guard of the outer loop, which is false, and sets $\mathit{turn} := 1$ just before the critical section.
Next, process 1 indicates it wants to access the critical section. Since $\mathit{flag}[0]$ is $\true$, process 1 enters the outer loop, and since $\mathit{turn} = 1$, it will not enter into the if-statement, so it will keep cycling here until $\mathit{flag}[0]$ becomes $\false$.
What we see here is that, because process 1 is continuously cycling through the outer loop, process 0 never gets
a chance to actually enter into its critical section.
This is a typical fairness issue.

We could try to alter the formula in such a way that unfair paths such as in the counterexample satisfy the property, and are thus, essentially, ignored.
In this case, we can do so by saying that each sequence not containing an \inlinemcrl{enter(i)} action ends in an infinite sequence of \inlinemcrl{get_flag} and \inlinemcrl{get_turn} actions.
This results in the following formula.

\begin{lstlisting}[language=mCRL2]
[true*]forall i:Nat.val(i<=1) => 
      [exists b:Bool . wish(i)|set_flag(i,b)] 
            nu X.mu Y.
                  ([!enter(i) && !(exists i1:Nat.get_flag(i1, true) || get_turn(i1))]Y &&
                    [exists i1:Nat.get_flag(i1, true) ||  get_turn(i1)]X)
\end{lstlisting}

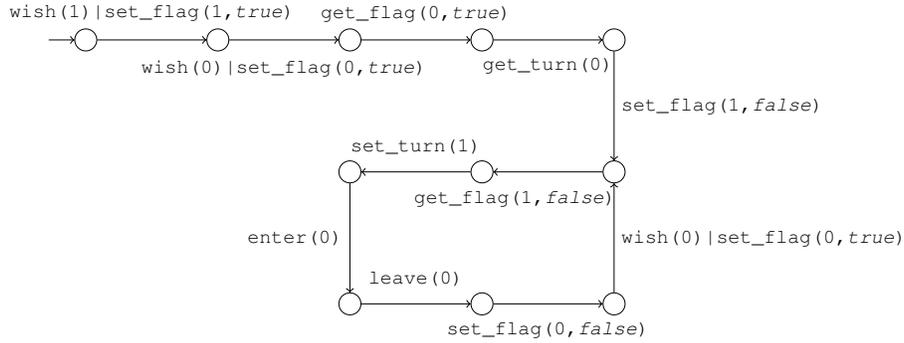
\begin{figure}[!ht]
  \centering
\resizebox{\textwidth}{!}{
  \begin{tikzpicture}[node distance=60pt]
    \lstset{language=mCRL2}
    \node[state, initial left] (s0) {};
    \node[state, right of=s0] (s1) {};
    \node[state, right of=s1] (s2) {};
    \node[state, right of=s2] (s3) {};
    \node[state, right of=s3] (s4) {};
    \node[state, below of=s4] (s5) {};
    \node[state, left of=s5] (s6) {};
    \node[state, left of=s6] (s7) {};
    \node[state, below of=s7] (s8) {};
    \node[state, right of=s8] (s9) {};
    \node[state, right of=s9] (s10) {};

    \draw[->]
      (s0) edge node[above,yshift=5pt] {\inlinemcrl{wish(1)|set_flag(1,true)}} (s1)
      (s1) edge node[below,yshift=-5pt] {\inlinemcrl{wish(0)|set_flag(0,true)}} (s2)
      (s2) edge node[above,yshift=5pt] {\inlinemcrl{get_flag(0,true)}} (s3)
      (s3) edge node[below,yshift=-5pt] {\inlinemcrl{get_turn(0)}} (s4)

      (s4) edge node[right] {\inlinemcrl{set_flag(1,false)}} (s5)
      (s5) edge node[below,yshift=-5pt,xshift=-15pt] {\inlinemcrl{get_flag(1,false)}} (s6)
      (s6) edge node[above,yshift=5pt] {\inlinemcrl{set_turn(1)}} (s7)
      (s7) edge node[left] {\inlinemcrl{enter(0)}} (s8)
      (s8) edge node[above,yshift=5pt] {\inlinemcrl{leave(0)}} (s9)
      (s9) edge node[below,yshift=-5pt] {\inlinemcrl{set_flag(0,false)}} (s10)
      (s10) edge node[right] {\inlinemcrl{wish(0)|set_flag(0,true)}} (s5)
      ;
  \end{tikzpicture}}
  \caption{Counterexample of the eventual access property under fairness for Dekker's algorithm.}
  \label{fig:dekker-eventual-access-if-fair-counterexample}
\end{figure}

Unfortunately, if we verify this property, we find it also does not hold.
We get a different counterexample, which is shown in Figure~\ref{fig:dekker-eventual-access-if-fair-counterexample}.

What we see is that after process 1 wishes to enter its critical section, process 0 can come and enter the critical section infinitely many times, preventing process 1 from entering the critical section.
A closer inspection reveals that this is because, to allow process 0 to enter, process 1 sets its flag to false, and then waits until process 0's flag becomes false.
However, again, since we do not have any fairness guarantees, after setting its flag to false, process 0 can immediately request access to its critical section again, before process 1 observes that the flag became false.

We could, of course, try to change the property once more to exclude also this unfair execution. However, instead we we change our focus to Peterson's mutual exclusion protocol, as it is simpler, and therefore easier to analyse.



\subsection{Peterson's mutual exclusion algorithm}

Some of the issues in Dekker's algorithm, particularly regarding eventual access, are alleviated by Peterson's mutual exclusion protocol~\cite{Pet1981}. We previously presented a model of this algorithm in~\cite{GKL+2020}. We describe this in Algorithm~\ref{alg:peterson}.

\begin{center}
\begin{tabular}{l}

\label{alg:peterson}
\textbf{Data:} Global variables $\mathit{flag}[0], \mathit{flag}[1] \colon \Bool$ and $\mathit{turn} \colon \Nat$\\
$\mathit{flag}[i] := \true$; \\
$\mathit{turn} := 1{-}i$; \\
\textbf{while} $flag[1{-}i] \wedge \mathit{turn} = 1{-}i$ \textbf{do}
  \texttt{/* Busy waiting */} \textbf{end}\\
\texttt{/* Critical section */}\\
$\mathit{flag}[i] := \false$;
\end{tabular}\vspace{2ex}\\
{\small\textbf{Algorithm 4:} Peterson's algorithm for process $i$.}
\end{center}

In Algorithm~4, the $\mathit{turn}$ variable is used differently from Dekker's algorithm.
When a process requests access to the critical section by setting its flag, it will behave politely, and let the other process go first.
It waits until either the other process does not ask for access to the critical section, i.e. $\mathit{flag}[1{-}i]$ is $\false$, or the other process arrived later, in which case $\mathit{turn} = i$.

\paragraph{Model.}
Peterson's algorithm can be modelled in mCRL2 using the same principles we have used before.
The structure of the initialization is completely analogous to that of the previous models.
A single process executing Peterson's algorithm can be modelled as follows.

\newpage
\begin{lstlisting}[linerange=34-41,language=mcrl2]
proc Process(i:Nat) =
               wish(i)|set_flag_s(i, true).
               set_turn_s(other(i)). 
               (get_flag_r(other(i), false) + get_turn_r(i)).
               enter(i).
               leave(i). 
               set_flag_s(i, false).
               Process(i);
\end{lstlisting}

Note that we use the fact that the negation of the guard of the loop is $\neg \mathit{flag}[1{-}i] \lor \mathit{turn} {=} i$, hence we can still use communicating actions to block until the guard becomes $\false$.

\paragraph{Verification.}
This model satisfies all properties we investigated so far, including eventual access.
This confirms the intuition we presented when introducing the algorithm.
Let us now switch our attention to bounded overtaking, which we have not investigated yet.

Bounded overtaking says that if one process indicates its wish to enter, other processes can at most
enter the critical section \inlinemcrl{B} times before this process is allowed to enter. It can be expressed
as follows.

\begin{lstlisting}[language=mcrl2]
[true*] forall i:Nat.[exists b:Bool.wish(i)|set_flag(i,b)]
                                         (nu Y(n:Nat = 0).val(n<=B) &&
                                                                           [!(exists i1:Nat.enter(i1))]Y(n) &&
                                                                           [enter(other(i))]Y(n+1) )
\end{lstlisting}

In this formula, for all processes \inlinemcrl{i}, whenever process \inlinemcrl{i} wishes to enter its critical section, we
start to count the number of times the other process enters its critical section using the parameter \inlinemcrl{n}.
All actions other than \inlinemcrl{enter} maintain the current value.
Meanwhile, the property asserts that \inlinemcrl{n<=B}, i.e., the bound is satisfied.

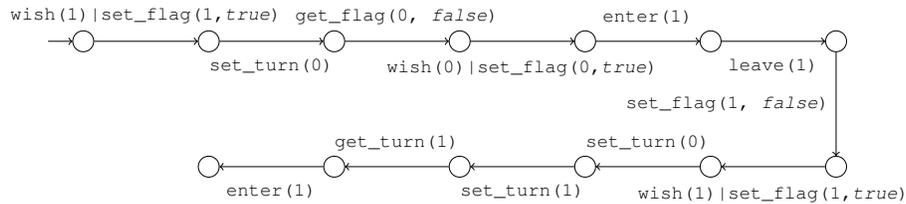
\begin{figure}
\centering
\resizebox{\textwidth}{!}{
\begin{tikzpicture}[node distance=60pt]
  \lstset{language=mCRL2}
  \node[state, initial left] (s0) {};
  \node[state, right of=s0] (s1) {};
  \node[state, right of=s1] (s2) {};
  \node[state, right of=s2] (s3) {};
  \node[state, right of=s3] (s4) {};
  \node[state, right of=s4] (s5) {};
  \node[state, right of=s5] (s6) {};
  \node[state, below of=s6] (s7) {};
  \node[state, left of=s7] (s8) {};
  \node[state, left of=s8] (s9) {};
  \node[state, left of=s9] (s10) {};
  \node[state, left of=s10] (s11) {};
  \node[state, left of=s11] (s12) {};

  \draw[->]
    (s0) edge node[above,yshift=5pt] {\inlinemcrl{wish(1)|set_flag(1,true)}} (s1)
    (s1) edge node[below,yshift=-5pt] {\inlinemcrl{set_turn(0)}} (s2)
    (s2) edge node[above,yshift=5pt] {\inlinemcrl{get_flag(0, false)}} (s3)
    (s3) edge node[below,yshift=-5pt] {\inlinemcrl{wish(0)|set_flag(0,true)}} (s4)
    (s4) edge node[above,yshift=5pt] {\inlinemcrl{enter(1)}} (s5)
    (s5) edge node[below,yshift=-5pt] {\inlinemcrl{leave(1)}} (s6)
    (s6) edge node[left] {\inlinemcrl{set_flag(1, false)}} (s7)
    (s7) edge node[below,yshift=-5pt] {\inlinemcrl{wish(1)|set_flag(1,true)}} (s8)
    (s8) edge node[above,yshift=5pt] {\inlinemcrl{set_turn(0)}} (s9)
    (s9) edge node[below,yshift=-5pt] {\inlinemcrl{set_turn(1)}} (s10)
    (s10) edge node[above,yshift=5pt] {\inlinemcrl{get_turn(1)}} (s11)
    (s11) edge node[below,yshift=-5pt] {\inlinemcrl{enter(1)}} (s12)
    ;
\end{tikzpicture}}

\caption{Counterexample: Peterson does not satisfy bounded overtaking for $B {=} 1$.}
\label{fig:peterson-bounded-overtaking-1-counterexample}
\end{figure}

Intuitively, we may expect that whenever a process wishes to enter its critical section, the other process may enter once first. However, if we check bounded overtaking with \inlinemcrl{B}$ {=} 1$, we get the counterexample shown in Figure~\ref{fig:peterson-bounded-overtaking-1-counterexample}.

Let us take a close look at the counterexample.
First, process 1 wishes to enter its critical section; it sets its flag, sets the turn to 0 and then checks the flag of process 0, which is currently $\false$.
At this point, process 1 is allowed to enter its critical section.
However, before entering, process 0 also wishes to enter its critical section, and sets its flag.
Subsequently, process 1 actually enters the critical section, sets the turn to process 0, and only then process 0 sets the turn to 1, ultimately allowing process 1 to enter a second time.
Hence, because process 0 is stalled after setting its flag, but before setting the turn to process 1, process 1 can overtake process 0 and enter a second time.

This leads to the question of whether overtaking for higher values of \inlinemcrl{B} is also possible.
By reverifying the formula for \inlinemcrl{B}$=2$, we find that the formula is valid. Bounded overtaking
for Peterson's mutual exclusion protocol is limited to at most $1$ times.

In~\cite{GKL+2020} we investigated a version of Peterson's algorithm where, initially, one of the flags is set to $\true$ instead of $\false$.
This alternative initialisation was, at some point, described on Wikipedia~\cite{PetersonWikipedia2015}.
It turns out that all four properties discussed above hold.
However, if the flag of process 1 is initially $\true$, process 0 will need the cooperation of process 1 to be allowed to enter for the first time, as it will set the turn to 1, and will block on the busy waiting loop.

This poses the question whether our properties are sufficient to cover the desired properties of
mutual exclusion protocols. In particular one might want to verify the property that a process can
always eventually request entry, without the other process having to perform any action. This is done
by the following formula, which distinguishes Peterson's algorithm with and without correct initialisation.
\begin{lstlisting}[language=mcrl2]
[true*]forall i:Nat.val(i<=1) =>
                    <!(wish(other(i))|set_flag(other(i), true)||
                         set_turn(i)||
                         get_flag(i, false)||
                         get_turn(other(i))||
                         enter(other(i))||
                         leave(other(i))||
                         set_flag(other(i),false))*><wish(i)|set_flag(i, true)>true
\end{lstlisting}
\section{Epilogue}
We went through several versions of mutual exclusion algorithms and showed that their correctness can
be formulated and investigated using modal formulas. Although it requires skill and experience to write down
process algebraic specifications, and in particular modal formulas with data, they provide a powerful pair of tools
to investigate and design protocols and distributed algorithms. We used it to study and design many systems
varying from games \cite{DBLP:journals/siamrev/GrooteWZ16,DBLP:conf/birthday/GrooteV17} to core protocols
for embedded systems \cite{DBLP:journals/lmcs/GrooteW20}.

When the systems that are modelled become more complex, the state space grows, and verification of modal
formulas becomes more time consuming, up to a point where the state space cannot be handled by contemporary
tools. It turns out that the style of modelling has a substantial influence on how complex systems can
become. In \cite{DBLP:journals/stvr/GrooteKO15} 7 different specification guidelines are presented to keep the state space
small.

\bibliographystyle{splncs04}
\bibliography{tutorial}

\begin{thebibliography}{10}
\providecommand{\url}[1]{\texttt{#1}}
\providecommand{\urlprefix}{URL }
\providecommand{\doi}[1]{https://doi.org/#1}

\bibitem{DBLP:journals/fac/AptO19}
Apt, K.R., Olderog, E.: Fifty years of hoare's logic. Formal Aspects Comput.
  \textbf{31}(6),  751--807 (2019). \doi{10.1007/s00165-019-00501-3},
  \url{https://doi.org/10.1007/s00165-019-00501-3}

\bibitem{DBLP:conf/icalp/BergstraK84}
Bergstra, J.A., Klop, J.W.: The algebra of recursively defined processes and
  the algebra of regular processes. In: Paredaens, J. (ed.) Automata, Languages
  and Programming, 11th Colloquium, Antwerp, Belgium, July 16-20, 1984,
  Proceedings. Lecture Notes in Computer Science, vol.~172, pp. 82--94.
  Springer (1984). \doi{10.1007/3-540-13345-3\_7},
  \url{https://doi.org/10.1007/3-540-13345-3\_7}

\bibitem{DBLP:series/txtcs/BertotC04}
Bertot, Y., Cast{\'{e}}ran, P.: Interactive Theorem Proving and Program
  Development - Coq'Art: The Calculus of Inductive Constructions. Texts in
  Theoretical Computer Science. An {EATCS} Series, Springer (2004).
  \doi{10.1007/978-3-662-07964-5},
  \url{https://doi.org/10.1007/978-3-662-07964-5}

\bibitem{DBLP:conf/fmics/BeusekomGHHWWW17}
van Beusekom, R., Groote, J.F., Hoogendijk, P.F., Howe, R., Wesselink, W.,
  Wieringa, R., Willemse, T.A.C.: Formalising the dezyne modelling language in
  mcrl2. In: Petrucci, L., Seceleanu, C., Cavalcanti, A. (eds.) Critical
  Systems: Formal Methods and Automated Verification - Joint 22nd International
  Workshop on Formal Methods for Industrial Critical Systems - and - 17th
  International Workshop on Automated Verification of Critical Systems,
  FMICS-AVoCS 2017, Turin, Italy, September 18-20, 2017, Proceedings. Lecture
  Notes in Computer Science, vol. 10471, pp. 217--233. Springer (2017).
  \doi{10.1007/978-3-319-67113-0\_14},
  \url{https://doi.org/10.1007/978-3-319-67113-0\_14}

\bibitem{DBLP:books/el/07/BradfieldS07}
Bradfield, J.C., Stirling, C.: Modal mu-calculi. In: Blackburn, P., van
  Benthem, J.F.A.K., Wolter, F. (eds.) Handbook of Modal Logic, Studies in
  logic and practical reasoning, vol.~3, pp. 721--756. North-Holland (2007).
  \doi{10.1016/s1570-2464(07)80015-2},
  \url{https://doi.org/10.1016/s1570-2464(07)80015-2}

\bibitem{DBLP:journals/scp/BrandG15}
van~den Brand, M., Groote, J.F.: Software engineering: Redundancy is key. Sci.
  Comput. Program.  \textbf{97},  75--81 (2015).
  \doi{10.1016/j.scico.2013.11.020},
  \url{https://doi.org/10.1016/j.scico.2013.11.020}

\bibitem{DBLP:journals/tcs/CranenGR11}
Cranen, S., Groote, J.F., Reniers, M.A.: A linear translation from ctl* to the
  first-order modal {\(\mu\)} -calculus. Theor. Comput. Sci.  \textbf{412}(28),
   3129--3139 (2011). \doi{10.1016/j.tcs.2011.02.034},
  \url{https://doi.org/10.1016/j.tcs.2011.02.034}

\bibitem{Dij1965}
Dijkstra, E.W.: Solution of a problem in concurrent programming control.
  Communications of the ACM  \textbf{8}(9), ~569 (Sep 1965).
  \doi{10.1145/365559.365617}

\bibitem{Dij1962}
Dijkstra, E.W.: Over de sequentialiteit van procesbeschrijvingen (Undated, 1962
  or 1963)

\bibitem{DBLP:journals/sttt/GaravelLMS13}
Garavel, H., Lang, F., Mateescu, R., Serwe, W.: {CADP} 2011: a toolbox for the
  construction and analysis of distributed processes. Int. J. Softw. Tools
  Technol. Transf.  \textbf{15}(2),  89--107 (2013).
  \doi{10.1007/s10009-012-0244-z},
  \url{https://doi.org/10.1007/s10009-012-0244-z}

\bibitem{DBLP:journals/sttt/Gibson-Robinson16}
Gibson{-}Robinson, T., Armstrong, P.J., Boulgakov, A., Roscoe, A.W.: {FDR3:} a
  parallel refinement checker for {CSP}. Int. J. Softw. Tools Technol. Transf.
  \textbf{18}(2),  149--167 (2016). \doi{10.1007/s10009-015-0377-y},
  \url{https://doi.org/10.1007/s10009-015-0377-y}

\bibitem{GKL+2020}
Groote, J.F., Keiren, J.J.A., Luttik, B., {de Vink}, E.P., Willemse, T.A.C.:
  Modelling and {{Analysing Software}} in {{mCRL2}}. In: Arbab, F., Jongmans,
  S.S. (eds.) Formal {{Aspects}} of {{Component Software}}. pp. 25--48. Lecture
  {{Notes}} in {{Computer Science}}, {Springer International Publishing},
  {Cham} (2020). \doi{10.1007/978-3-030-40914-2\_2}

\bibitem{DBLP:journals/stvr/GrooteKO15}
Groote, J.F., Kouters, T.W.D.M., Osaiweran, A.: Specification guidelines to
  avoid the state space explosion problem. Softw. Test. Verification Reliab.
  \textbf{25}(1),  4--33 (2015). \doi{10.1002/stvr.1536},
  \url{https://doi.org/10.1002/stvr.1536}

\bibitem{DBLP:books/mit/GrooteM2014}
Groote, J.F., Mousavi, M.R.: Modeling and Analysis of Communicating Systems.
  {MIT} Press (2014),
  \url{https://mitpress.mit.edu/books/modeling-and-analysis-communicating-systems}

\bibitem{DBLP:conf/birthday/GrooteV17}
Groote, J.F., de~Vink, E.P.: Problem solving using process algebra considered
  insightful. In: Katoen, J., Langerak, R., Rensink, A. (eds.) ModelEd, TestEd,
  TrustEd - Essays Dedicated to Ed Brinksma on the Occasion of His 60th
  Birthday. Lecture Notes in Computer Science, vol. 10500, pp. 48--63. Springer
  (2017). \doi{10.1007/978-3-319-68270-9\_3},
  \url{https://doi.org/10.1007/978-3-319-68270-9\_3}

\bibitem{DBLP:journals/siamrev/GrooteWZ16}
Groote, J.F., Wiedijk, F., Zantema, H.: A probabilistic analysis of the game of
  the goose. {SIAM} Rev.  \textbf{58}(1),  143--155 (2016).
  \doi{10.1137/140983781}, \url{https://doi.org/10.1137/140983781}

\bibitem{DBLP:journals/lmcs/GrooteW20}
Groote, J.F., Willemse, T.A.C.: A symmetric protocol to establish service level
  agreements. Log. Methods Comput. Sci.  \textbf{16}(3) (2020),
  \url{https://lmcs.episciences.org/6812}

\bibitem{DBLP:journals/jacm/HennessyM85}
Hennessy, M., Milner, R.: Algebraic laws for nondeterminism and concurrency. J.
  {ACM}  \textbf{32}(1),  137--161 (1985). \doi{10.1145/2455.2460},
  \url{https://doi.org/10.1145/2455.2460}

\bibitem{DBLP:journals/entcs/HopcroftB05}
Hopcroft, P.J., Broadfoot, G.H.: Combining the box structure development method
  and {CSP} for software development. Electron. Notes Theor. Comput. Sci.
  \textbf{128}(6),  127--144 (2005). \doi{10.1016/j.entcs.2005.04.008},
  \url{https://doi.org/10.1016/j.entcs.2005.04.008}

\bibitem{DBLP:journals/corr/LeinoW14}
Leino, K.R.M., W{\"{u}}stholz, V.: The dafny integrated development
  environment. In: Dubois, C., Giannakopoulou, D., M{\'{e}}ry, D. (eds.)
  Proceedings 1st Workshop on Formal Integrated Development Environment,
  {F-IDE} 2014, Grenoble, France, April 6, 2014. {EPTCS}, vol.~149, pp. 3--15
  (2014). \doi{10.4204/EPTCS.149.2}, \url{https://doi.org/10.4204/EPTCS.149.2}

\bibitem{DBLP:books/daglib/0067019}
Milner, R.: Communication and concurrency. {PHI} Series in computer science,
  Prentice Hall (1989)

\bibitem{DBLP:books/sp/NipkowPW02}
Nipkow, T., Paulson, L.C., Wenzel, M.: Isabelle/HOL - {A} Proof Assistant for
  Higher-Order Logic, Lecture Notes in Computer Science, vol.~2283. Springer
  (2002). \doi{10.1007/3-540-45949-9},
  \url{https://doi.org/10.1007/3-540-45949-9}

\bibitem{DBLP:journals/ese/OsaiweranSH14}
Osaiweran, A., Schuts, M., Hooman, J.: Experiences with incorporating formal
  techniques into industrial practice. Empir. Softw. Eng.  \textbf{19}(4),
  1169--1194 (2014). \doi{10.1007/s10664-013-9251-2},
  \url{https://doi.org/10.1007/s10664-013-9251-2}

\bibitem{PetersonWikipedia2015}
Peterson's algorithm.
  \url{https://en.wikipedia.org/wiki/Peterson%27s_algorithm} (2015), accessed
  May 17

\bibitem{Pet1981}
Peterson, G.L.: Myths about the mutual exclusion problem. Information
  Processing Letters  \textbf{12}(3),  115--116 (Jun 1981).
  \doi{10.1016/0020-0190(81)90106-X}

\bibitem{DBLP:conf/cade/WesselinkW18}
Wesselink, W., Willemse, T.A.C.: Evidence extraction from parameterised boolean
  equation systems. In: Benzm{\"{u}}ller, C., Otten, J. (eds.) Proceedings of
  the 3rd International Workshop on Automated Reasoning in Quantified
  Non-Classical Logics {(ARQNL} 2018) affiliated with the International Joint
  Conference on Automated Reasoning {(IJCAR} 2018), Oxford, UK, July 18, 2018.
  {CEUR} Workshop Proceedings, vol.~2095, pp. 86--100. CEUR-WS.org (2018),
  \url{http://ceur-ws.org/Vol-2095/paper6.pdf}

\end{thebibliography}

\end{document}